\makeatletter 

\documentclass[aps,twocolumn,pra,tightenlines,floatfix,showpacs,superscriptaddress]{revtex4-2}

\usepackage{graphicx}
\usepackage{epstopdf}
\usepackage[english]{babel}
\usepackage{amsmath}
\allowdisplaybreaks
\usepackage{mathtools}
\usepackage{ntheorem}
\theoremseparator{:}
\theoremheaderfont {\it}

\usepackage{amssymb}
\usepackage{times}
\usepackage{appendix}
\usepackage{bm}
\usepackage{float}
\usepackage{color}
\usepackage{url}
\usepackage[usenames,dvipsnames]{xcolor}

\usepackage{caption}
\usepackage[percent]{overpic}
\usepackage{indentfirst} 

\newcommand{\Rmnum}[1]{\expandafter\@slowromancap\romannumeral #1@}

\newcommand{\PT}{\mathcal{PT}}
\usepackage{array}
\usepackage{booktabs}
\makeatother 
\usepackage[colorlinks=true,linkcolor=Blue,urlcolor=Blue,citecolor=Blue,anchorcolor=Red]{hyperref}
\usepackage{cleveref}
\crefname{equation}{Eq.~}{Eq.~}
\crefname{figure}{Fig.~}{Fig.~}
\usepackage{natbib}

\begin{document}
		\title{Magnetometry and slow light in $\PT$-symmetric cavity optomechanics}

	\author{Yuguo Su}
	
	
	\author{Tianshun Tan}
	
	\author{Yiying Yan}
	
	\author{Hengyan Wang}
	\affiliation{School of Science, Zhejiang University of Science and Technology, Hangzhou 310023, China}

	\author{Elijah Omollo Ayieta}
	\affiliation{Department of Physics, University of Nairobi, Nairobi 30197, Kenya}
	
	\author{Szabolcs Kelemen}
	\affiliation{Department of Physics, Babe\c{s}-Bolyai University, 400084 Cluj-Napoca, Romania}

	\author{Slobodan Rado\v{s}evi\'c}
	\affiliation{Department of Physics, Faculty of Science, University of Novi Sad, Trg Dositeja Obradovi\'ca 4, 21000 Novi Sad, Serbia}

	\author{JunYan Luo}
	\affiliation{School of Science, Zhejiang University of Science and Technology, Hangzhou 310023, China}
	
	\author{Yi-Xiao Huang}
	\affiliation{School of Science, Zhejiang University of Science and Technology, Hangzhou 310023, China}
	\author{Xingyu Zhang}
	\email{zhangxingyu49@zust.edu.cn}
	\affiliation{School of Science, Zhejiang University of Science and Technology, Hangzhou 310023, China}
	
	\begin{abstract}
		The generation of optical higher-order sidebands, which produces new frequencies with regular spectral separation, holds significant promise for applications in optical communication and quantum metrology.
		We investigate second-order sideband generation and tunable slow-fast light effects in a parity-time ($\PT$)-symmetric cavity optomechanics and propose an enhanced magnetometer scheme.
		We demonstrate that the $\PT$-symmetric structure, an external magnetic field, and the photon tunneling strength provide powerful and flexible control over both the second-order sideband efficiency, as well as the broadening and switching of slow-fast light propagation. 
		Operating in the $\PT$-symmetry-protected phases, our magnetometer based on the second-order sideband achieves a theoretical sensitivity at the picotesla level, surpassing the performance of previous optomechanical approaches.
		Our results offer new strategies for magnetic field sensing and optical transmission control, paving the way toward applications in quantum sensing and optical information storage.
	\end{abstract}
	
	\maketitle
	
	\section{Introduction}\label{Sec.I}
		Optomechanical systems have become versatile theoretical and experimental platforms, enabled by advances in nanofabrication~\cite{biswas_advances_2012,li_recent_2021,chen_nanofabrication_2015}, for studying the interplay of optics, mechanics, and electromagnetism at the microscale.
		As an innovative platforms that couples optical and mechanical modes via radiation pressure, it has enabled the observation of a range of distinctive physical phenomena, such as optomechanically induced transparency (OMIT)~\cite{agarwal_electromagnetically_2010,lai_tunable_2020}, high-order sideband generation~\cite{jiao_optomechanical_2018,suzuki_nonlinear_2015,xiao_tunable_2020}, and slow- and fast-light effects~\cite{liao_transparency_2020,safavi-naeini_electromagnetically_2011,singh_tunable_2022}.
		Owing to its high degree of tunability and control, optomechanical interactions represent an essential resource for numerous cutting-edge fields, including enhanced quantum sensing~\cite{xiong_precision_2017, zhang_precision_2012,wang_precision_2015,li_nonlinear_2012,wang_precision_2017,su_quantum_2024,su_dynamical_2025,yuguosu_parametrized_2021,su_approaching_2025}, macroscopic quantum phenomena~\cite{delić_cooling_2020,ma_nonclassical_2021,lei_quantum_2023,su_spin_2020b}, and quantum computing~\cite{qiu_laser_2020b,wen_groundstate_2023,fiaschi_optomechanical_2021,teh_simulation_2017}.
		
		As a typical nonlinear interaction phenomenon, high-order sideband generation has drawn considerable interest in fields such as optical communications~\cite{xiong_carrierenvelope_2013}, optical frequency combs~\cite{li_generation_2018}, and high-sensitivity metrology~\cite{hodaei_enhanced_2017}.
		The generation of optical high-order sidebands depends critically on the input light, and these nonlinear effects typically become observable only under strong driving fields or specially tailored driving pulses.
		Compared with first-order sidebands, the high-order sideband signal is considerably weaker, and its faint nonlinear response poses a notable detection challenge, which has hindered its broader practical utilization. 
		Consequently, improving the generation efficiency of high-order sidebands has become a topic of active research.
		Several schemes have been proposed to enhance the high-order sideband effect, including the optomechanical systems using nonlinear materials~\cite{jiao_optomechanical_2018,li_enhancement_2019,liao_generation_2023}, hybrid electromagnetic optomechanical systems ~\cite{zhang_pt_2020,liu_precision_2018,yu_modelling_2018,wu_nanocavity_2017}, and parity-time ($\PT$)-symmetric cavity optomechanics (COM) ~\cite{liu_metrology_2016,jiao_nonlinear_2016}.
		The distinctive behavior of $\mathcal{PT}$-symmetric systems near their phase transition point, combined with the high sensitivity of optomechanical high-order sidebands to perturbations such as magnetic fields, makes such systems a promising platform for magnetic field sensing~\cite{liu_metrology_2016}.
		Experimentally, nanoscale cavity optomechanical devices have achieved a peak magnetic sensitivity of $400$~nT~\cite{forstner_cavity_2012,forstner_ultrasensitive_2014}, with theoretical studies indicating that $\mathcal{PT}$-symmetric systems operating with first-order sidebands could push sensitivities to $0.01$~nT level~\cite{zhang_pt_2020}.
		
		Tunable slow- and fast-light effects, arising from the modification of the optical group velocity via optomechanical interaction, have been experimentally demonstrated~\cite{zhao_new_2009}.
		Slow light, in particular, has enabled a wide range of applications, including storing light~\cite{fiaschi_optomechanical_2021,he_light_2016}, implementing buffers for tunable optical semiconductor nanostructures~\cite{chang-hasnain_variable_2003}, improving the performance of optical instruments~\cite{peng_rotation_2007,zou_effect_2006,thévenaz_time_2006,liao_transparency_2020}.
		By enhancing light-matter interaction, it provides improved control over the spectral bandwidth and allows light to be delayed or temporarily stored in all‑optical memories~\cite{krauss_why_2008}.
		This storage capability is essential for advancing technologies in quantum communication~\cite{fiaschi_optomechanical_2021}, quantum computation~\cite{teh_simulation_2017}, and quantum sensing~\cite{krauss_why_2008,merklein_brillouinbased_2018}.
		Currently, the storage of light pulses has been experimentally realized in diverse platforms such as photonic crystals~\cite{krauss_slow_2007}, Bragg gratings~\cite{mok_dispersionless_2006}, Bose-Einstein condensates~\cite{hau_light_1999}, and optomechanical systems~\cite{safavi-naeini_electromagnetically_2011}.
		However, the integration of magnetomechanical interactions and $\PT$-symmetric structures into optomechanical systems for magnetic field sensing and slow-fast light control remains relatively unexplored~\cite{liao_generation_2023,jiao_optomechanical_2018}.
		
		In this work, an enhanced magnetometer scheme is proposed, and the second-order sideband generation and tunable slow-fast light effects are investigated in $\PT$-symmetric COM.
		We show that the combined effect of the $\mathcal{PT}$-symmetric structure, an external magnetic field, and the photon tunneling strength can significantly enhance both the second-order sideband efficiency and the optical group delay.
		Moreover, switching between slow‑ and fast‑light regimes can be achieved by tuning the driving power.
		We further propose an enhanced magnetometer scheme based on the second-order sidebands, which exhibits a theoretical sensitivity at the picotesla level when operated in the $\mathcal{PT}$-symmetry‑protected phase.
		This magnetometer offers several advantages, including high measurement accuracy, a low driving threshold, ultra‑low mass, a high mechanical quality factor, and excellent integrability.
		Our findings advance the understanding of tunable light-matter interactions in COM and provide novel strategies for high‑precision magnetic‑field sensing and optical information storage.
		
		This paper is organized as follows.
		In Sec.~\ref{Sec.II}, the Hamiltonian of $\PT$-symmetric COM is introduced, and the efficiencies of the first-order and second-order sidebands, as well as their group delays, are derived.
		In Sec.~\ref{Sec.III}, we analyze the enhancement of second-order sideband generation.
		Subsequently, the precision and advantages of magnetic field detection are investigated by employing the enhanced second-order sidebands in $\PT$-symmetric COM.
		Finally, we discuss the enhancement of group delay and the switching of slow-fast light effects with driving power.
		Sec.~\ref{Sec.IV} summarizes our findings.
		\begin{figure}
			\centering
			\begin{minipage}{0.49\linewidth}
				\centering
				\begin{overpic}[width=\linewidth]{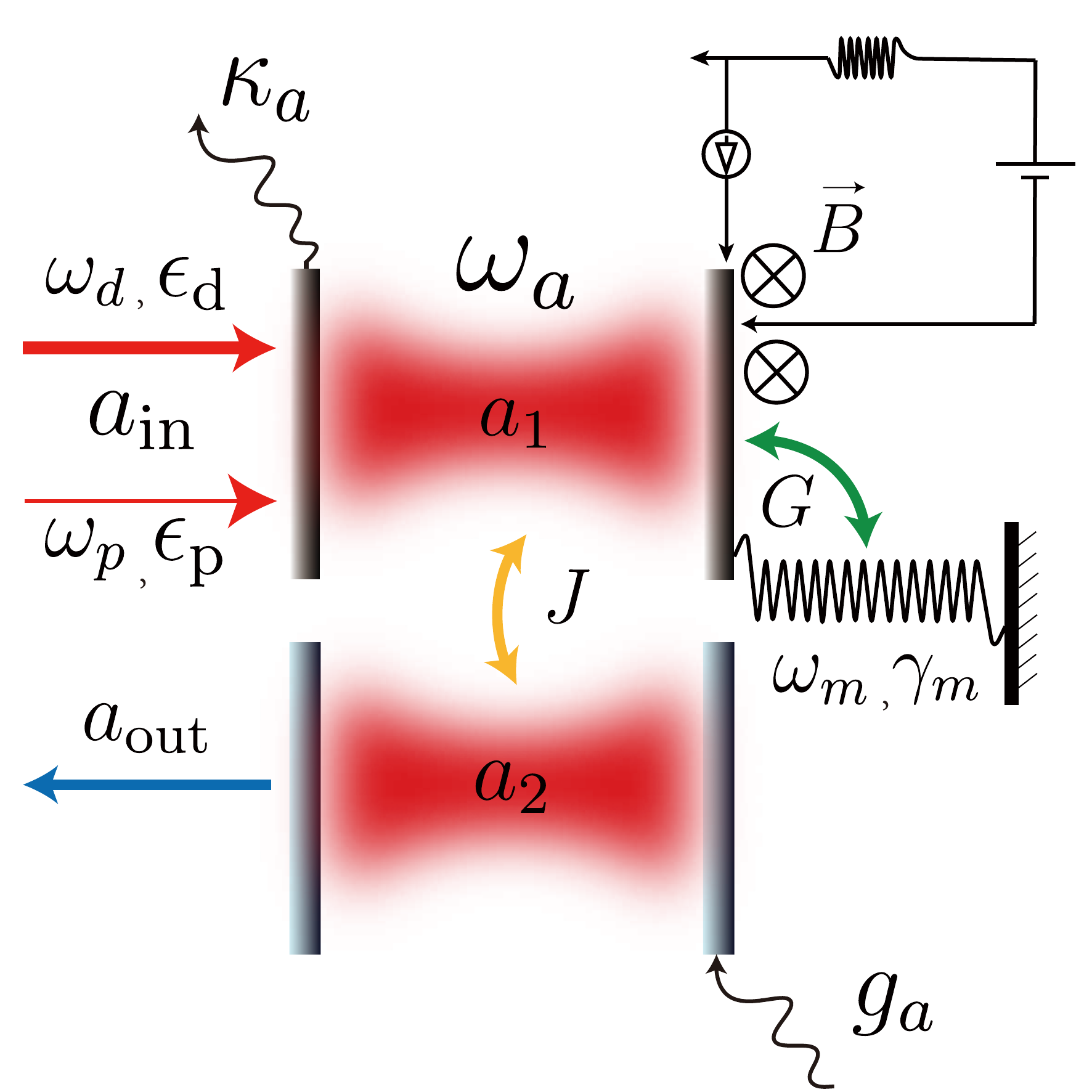}
					\put(0,90){(a)}
				\end{overpic}
			\end{minipage}
			\begin{minipage}{0.49\linewidth}
				\centering
				\begin{overpic}[width=\linewidth]{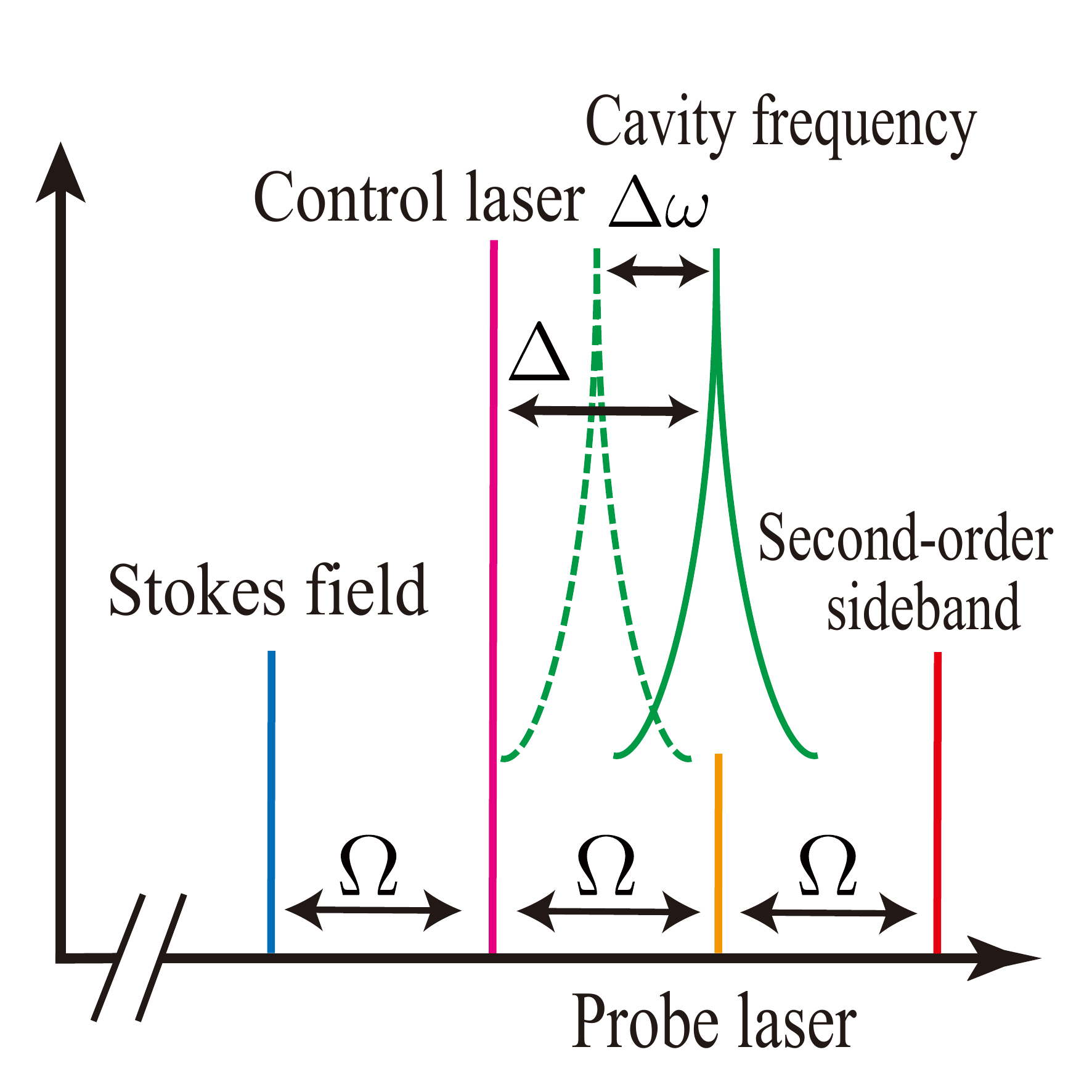}
					\put(0,90){(b)}
				\end{overpic}
			\end{minipage}\\
			\caption{Schematic illustration and frequency spectrum diagram of the $\PT$-symmetric COM. 
				(a) A lossy cavity is coupled to another cavity via photon tunneling.
				One mirror of the cavity field (which can be regarded as an electromechanical oscillator) interacts with the lossy cavity under a magnetic field of strength $B$, generated by a current intensity $I$. 
				The lossy cavity is driven by a strong driving field with frequency $\omega_{\text{d}}$ and amplitude $\epsilon_{\text{d}}$, and probed by a weak probe field with frequency $\omega_{\text{p}}$ and amplitude $\epsilon_{\text{p}}$.
				(b) The cavity mode experiences redshifted due to electromagnetic interaction.
			}\label{Fig1}
		\end{figure}
	
	\section{Theoretical model and dynamical equations}\label{Sec.II}
		A $\PT$-symmetric COM is depicted in Fig.~\ref{Fig1}, with its Hamiltonian given by
		\begin{align}
			H=&-\hbar\Delta_{a}\left(a_{1}^{\dagger}a_{1}+a_{2}^{\dagger}a_{2}\right)+\frac{p^{2}}{2m}+\frac{m}{2}\omega_{m}^{2}x^{2}\nonumber\\&+\hbar J\left(a_{1}^{\dagger}a_{2}+a_{2}^{\dagger}a_{1}\right)+\hbar Ga_{1}^{\dagger}a_{1}x+\zeta Bx\nonumber\\&+\mathrm{i}\hbar\sqrt{\eta_{c}\kappa_{a}}\left[\left(\epsilon_{d}a_{1}^{\dagger}+\epsilon_{p}a_{1}^{\dagger}e^{-\mathrm{i}\Omega t}\right)-\text{H.c.}\right], \label{H}
		\end{align}
		where the frequencies $\Delta_{a}=\omega_{\text{d}}-\omega_{a}$ and $\Omega=\omega_{\text{p}}-\omega_{\text{d}}$ are the pump-cavity and probe-pump detunings, respectively.
		Here, $\omega_{a}$ is cavity resonance frequency, $a_{1}^{\dagger}$ ($a_{1}$) and $a_{2}^{\dagger}$ ($a_{2}$) are creation (annihilation) operators of loss cavity and gain cavity.
		The mechanical oscillator has the momentum operator $p$, position operator $x$, resonance frequency $\omega_{m}$, and mass $m$.
		The term $\hbar J(a_{1}^{\dagger}a_{2}+a_{2}^{\dagger}a_{1})$ describes the coupling between the loss and gain cavities with strength $J$.
		Optomechanical interaction between the loss cavity and the mechanical oscillator is represented by $\hbar Ga_{1}^{\dagger}a_{1}x$, where $G$ is the corresponding coupling strength.
		The parameter $\zeta$ quantifies the electromagnetic interaction strength, which is proportional to the current intensity $I$ and the effective interaction range.
		The last term models the driving of the loss cavity by external input fields, with $\eta_{c}=1/2$ being the critical coupling parameter and $\kappa_{a}$ the loss rate of the loss cavity.
		The amplitudes $\epsilon_{\text{p},\text{d}}=\sqrt{P_{\text{p},\text{d}}/\hbar\omega_{\text{p},\text{d}}}$ are defined via the power $P_{\text{p},\text{d}}$ and frequency $\omega_{\text{p},\text{d}}$ of the input probe and pump fields.
		
		According to the Hamiltonian~\eqref{H}, the quantum dynamics of the system can be described by the following Heisenberg-Langevin equations~\cite{weis_optomechanically_2010,gardiner_input_1985,giovannetti_phasenoise_2001}:
		\begin{align}
			\dot{a}_{1}\!=&\left(\mathrm{i}\Delta_{a}-\mathrm{i}Gx-\frac{\kappa_{a}}{2}\right)a_{1}-\mathrm{i}Ja_{2}+\nonumber\\
			&\sqrt{\eta_{c}\kappa_{a}}\left(\epsilon_{\text{d}}+\epsilon_{\text{p}}e^{-\mathrm{i}\Omega t}\right),\\
			\dot{a}_{2}\!=&\left(\mathrm{i}\Delta_{a}+\frac{g_{a}}{2}\right)a_{2}-\mathrm{i}Ja_{1},\\
			\dot{p}=&-m\omega_{m}^{2}x-\hbar Ga_{1}^{\dagger}a_{1}-\zeta B-\gamma_{m}p,\\
			\dot{x}=&\frac{p}{m},\\
			\ddot{x}=&-\gamma_{m}\dot{x}-\omega_{m}^{2}x-\frac{1}{m}\left(\hbar Ga_{1}^{\dagger}a_{1}+\zeta B\right).
		\end{align}
		where $g_{a}$ is the gain rate of the gain cavity, and $\gamma_{m}$ is the damping rate of the mechanical oscillator.
		Since the probe field is much weaker than the driving field, the above equations can be solved using perturbation theory.
		Writing each operator as $O=\bar{O}+\delta O$, with $\bar{O}$ and $\delta O$ being the steady-state values and the corresponding quantum fluctuation, one obtains the steady-state solutions:
		
		\begin{align}
			&\bar{a}_{1} = -\frac{\sqrt{\eta_{c}\kappa_{a}}\epsilon_{\text{d}}}{\mathrm{i}\Delta_{a}-\mathrm{i}G\bar{x}-\frac{\kappa_{a}}{2}+\frac{J^{2}}{\mathrm{i}\Delta_{a}+\frac{g_{a}}{2}}}, &
			&\bar{a}_{2} = \frac{\mathrm{i}J\bar{a}_{1}}{\mathrm{i}\Delta_{a}+\frac{g_{a}}{2}}, \nonumber\\
			&\bar{x} = -\frac{\hbar G\left|\bar{a}_{1}\right|^{2}+\zeta B}{m\omega_{m}^{2}}, &
			&\bar{p} = 0.
		\end{align}
		The denominator $\mathrm{i}\Delta_{a}-\mathrm{i}G\bar{x}-\kappa_{a}/2+J^{2}/(\mathrm{i}\Delta_{a}+g_{a}/2)$ reflects the interplay between the driving detuning $\Delta_{a}$ and the intercavity coupling $J$, arising from the combined optomechanical and double‑cavity interactions present in the system.
		
		The evolution of the fluctuation terms can be derived as
		\begin{align}
			\frac{d\delta a_{1}}{dt}&=\left(\mathrm{i}\Delta-\frac{\kappa_{a}}{2}\right)\delta a_{1}-\mathrm{i}J\delta a_{2}\nonumber\\
			&\quad -\mathrm{i}G\left(\bar{a}_{1}\delta x+\delta a_{1}\delta x\right)+\sqrt{\eta_{c}\kappa_{a}}\epsilon_{\text{p}}e^{-\mathrm{i}\Omega t},\\
			\frac{d\delta a_{2}}{dt}&\!=\!\left(\mathrm{i}\Delta_{a}+\frac{g_{a}}{2}\right)\delta a_{2}-\mathrm{i}J\delta a_{1},\\
			\frac{d^{2}\delta x}{dt^{2}}&=-\gamma_{m}\frac{d\delta x}{dt}-\omega_{m}^{2}\delta x\nonumber\\
			&\quad-\frac{\hbar G}{m}\left(\bar{a}_{1}\delta a_{1}^{*}+\bar{a}_{1}^{*}\delta a_{1}+\delta a_{1}^{*}\delta a_{1}\right),
		\end{align}		
		with the effective detuning $\Delta=\Delta_{a}-G\bar{x}$ resulting from the coupling of the lossy cavity to the mechanical oscillator.
		The nonlinear terms $-\mathrm{i}G\delta a_{1}\delta x$ and $-(\hbar G/m)\delta a_{1}^{*}\delta a_{1}$ are responsible for generating output fields at frequencies $\omega_{\text{d}}\pm n\Omega$ with an integer $n$.
		Neglecting these terms would break the self-consistency of the Ansatz for the higher-order sidebands.
		In the present work, we restrict our analysis to the second-order sideband and disregard contributions from orders $n>2$.
		Consequently, we adopt the simplified Ans\"atze
		\begin{align}
			\delta a_{1}=&A_{1}^{-}e^{-\mathrm{i}\Omega t}+A_{1}^{+}e^{\mathrm{i}\Omega t}+A_{2}^{-}e^{-2\mathrm{i}\Omega t}+A_{2}^{+}e^{2\mathrm{i}\Omega t},\\
			\delta a_{2}=&B_{1}^{-}e^{-\mathrm{i}\Omega t}+B_{1}^{+}e^{\mathrm{i}\Omega t}+B_{2}^{-}e^{-2\mathrm{i}\Omega t}+B_{2}^{+}e^{2\mathrm{i}\Omega t},\\
			\delta x=&X_{1}e^{-\mathrm{i}\Omega t}+X_{1}^{*}e^{\mathrm{i}\Omega t}+X_{2}e^{-2\mathrm{i}\Omega t}+X_{2}^{*}e^{2\mathrm{i}\Omega t},
		\end{align}
		where $A_{n}^{\pm}$ and $B_{n}^{\pm}$ denote the complex amplitudes of the $n$-order upper ($-$) and lower ($+$) optical sidebands for loss and gain cavities, while $X_{n}$ represents the mechanical oscillation amplitudes at frequencies $n\Omega$.
		
		Solving the evolution equations via the above Ans\"atze, the amplitudes of the first-order upper (lower) sideband and the second-order upper sideband can be described by (see more details in Appendix):
		\begin{align}
			A_{1}^{-}=&\frac{\mathrm{i}\sqrt{\eta_{c}\kappa_{a}}\epsilon_{\text{p}}\left[h_{2}\left(\Omega\right)+h_{3}\left(\Omega\right)f\left(\Omega\right)\right]}{h_{1}\left(\Omega\right)},\label{A-1}\\
			(\!A_{1}^{+}\!)\!^{*}=&\frac{-\mathrm{i}\hbar G^{2}\bar{a}_{1}^{*2}\sqrt{\eta_{c}\kappa_{a}}\epsilon_{\text{p}}\chi\left(\Omega\right)h_{3}\left(\Omega\right)}{h_{1}\left(\Omega\right)},\label{A+1}\\
			A_{2}^{-}=&\!\frac{Gh_{2}\!\left(2\Omega\right)\!A_{1}^{-}\!X_{1}\lambda_{1}\!(\Omega)\!-\!\lambda_{2}\!h_{3}^{2}\!\left(2\Omega\right)\!\chi\left(2\Omega\right)X_{1}^{2}}{h_{1}\left(2\Omega\right)h_{2}\left(2\Omega\right)},\label{A-2}
		\end{align}		
		where the coefficient functions are $h_{1}(\Omega)=[\text{\ensuremath{\Delta_{a}}}^{2}-(\mathrm{i}g_{a}/2-\Omega)^{2}][\Delta^{2}-(\mathrm{i}\kappa_{a}/2+\Omega)^{2}+2\Delta f(\Omega)]-2J^{2}[\Delta_{a}f(\Omega)+\Delta\Delta_{a}-(\mathrm{i}g_{a}/2-\Omega)(\mathrm{i}\kappa_{a}/2+\Omega)]+J^{4}$, $h_{2}(\Omega)=(\Delta_{a}-\mathrm{i}g_{a}/2+\Omega)[(\Delta_{a}+\mathrm{i}g_{a}/2-\Omega)(\Delta-\mathrm{i}\kappa_{a}/2-\Omega)-J^{2}]$, $h_{3}(\Omega)=(\Delta_{a}-\mathrm{i}g_{a}/2+\Omega)(\Delta_{a}+\mathrm{i}g_{a}/2-\Omega)$, $f(\Omega)=\hbar G^{2}|\bar{a}_{1}|^{2}\chi(\Omega)$, $\chi(\Omega)=1/[m(\omega^{2}-\Omega^{2}-\mathrm{i}\gamma_{m}\Omega))]$,  $\lambda_{1}(2\Omega)=[f(2\Omega)h_{3}(2\Omega)+h_{2}(2\Omega)]-\hbar\bar{a}_{1}^{2}G^{2}h_{3}(2\Omega)\chi(2\Omega)$, $\lambda_{2}=\hbar G^{4}\left|\bar{a}_{1}\right|^{2}\bar{a}_{1}^{*}$.
		
		\begin{figure}
			\centering
			\begin{minipage}{0.49\linewidth}
				\centering				
				\begin{overpic}[width=\linewidth]{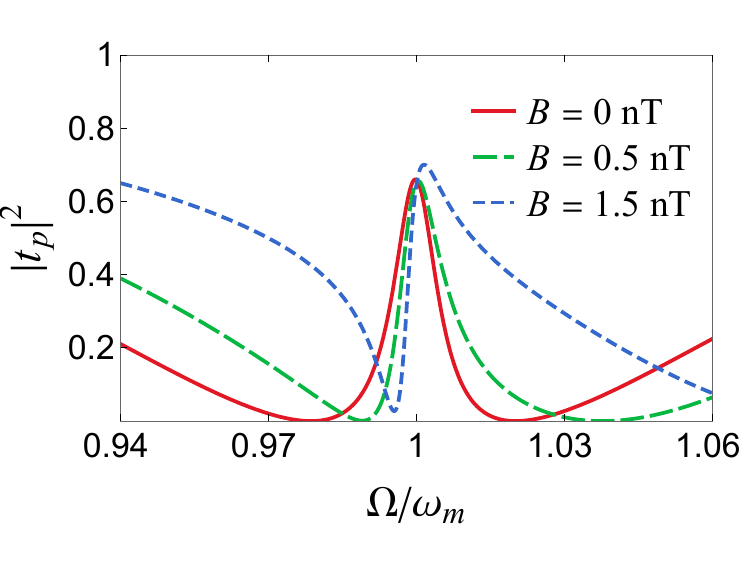}
					\put(0,73){(a)}
				\end{overpic}
			\end{minipage}
			\begin{minipage}{0.49\linewidth}
				\centering
				\begin{overpic}[width=\linewidth]{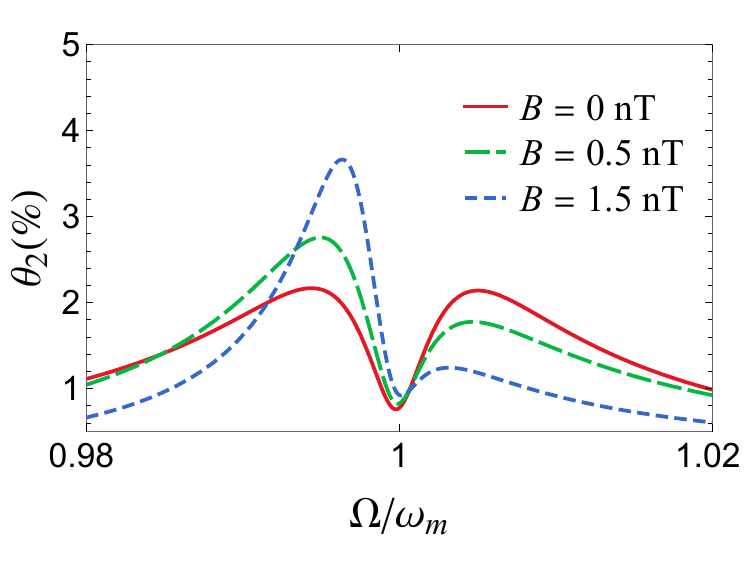}
					\put(0,73){(b)}
				\end{overpic}
			\end{minipage}\\
			\begin{minipage}{0.49\linewidth}
				\centering
				\begin{overpic}[width=\linewidth]{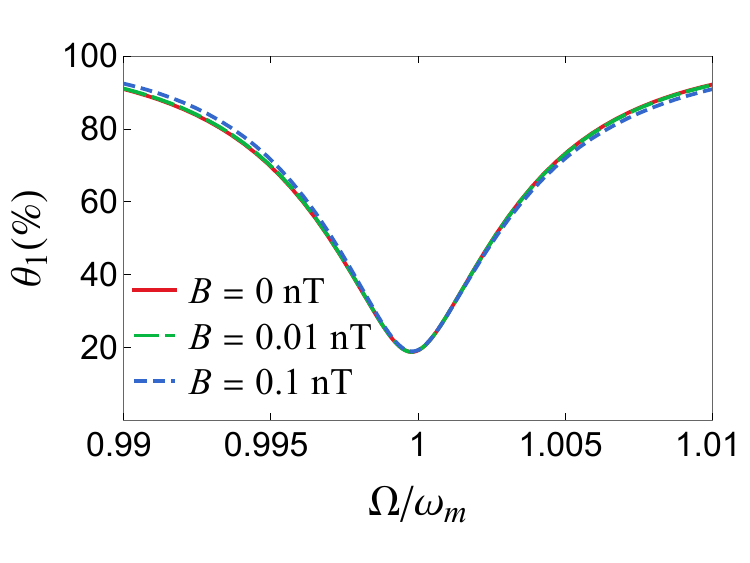}
					\put(0,73){(c)}
				\end{overpic}
			\end{minipage}
			\begin{minipage}{0.49\linewidth}
				\centering
				\begin{overpic}[width=\linewidth]{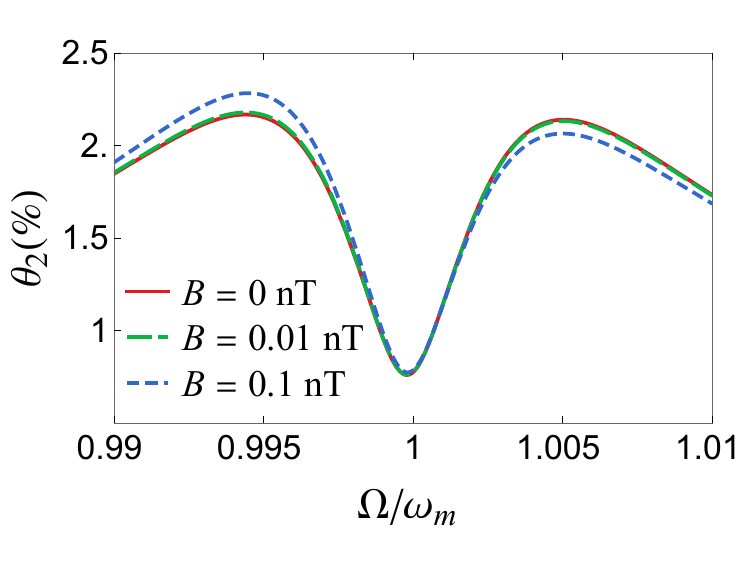}
					\put(0,73){(d)}
				\end{overpic}
			\end{minipage}\\
			\begin{minipage}{0.49\linewidth}
				\centering
				\begin{overpic}[width=\linewidth]{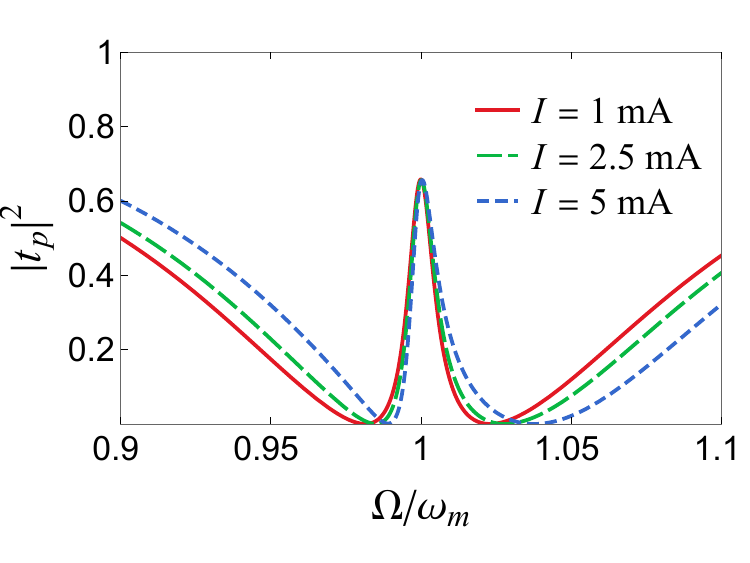}
					\put(0,73){(e)}
				\end{overpic}
			\end{minipage}
			\begin{minipage}{0.49\linewidth}
				\centering
				\begin{overpic}[width=\linewidth]{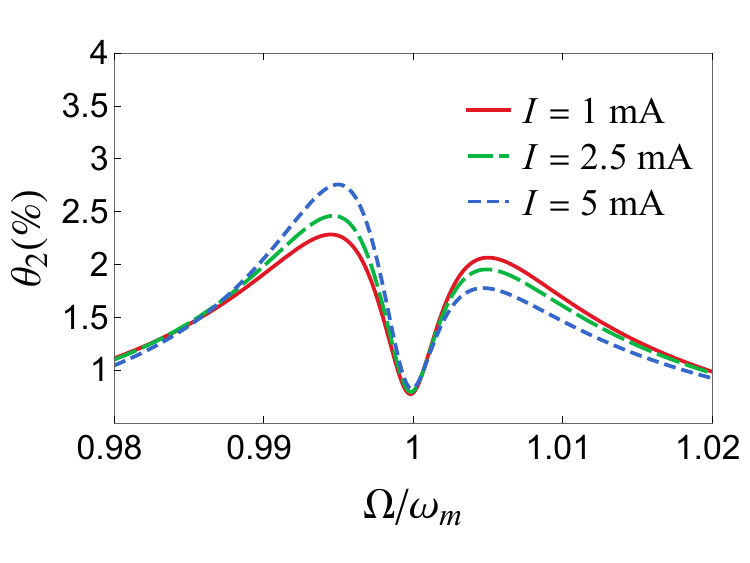}
					\put(0,73){(f)}
				\end{overpic}
			\end{minipage}
			\caption{
				Variations of the probe field transmission $|t_{p}|^{2}$, first-order upper sideband efficiency $\theta_{1}$, and second-order upper sideband efficiency $\theta_{2}$ with normalized detuning $\Omega/\omega_{m}$ under various magnetic field and current intensity settings.
				(a), (b) The magnetic field is set to $B=0$ nT, $0.5$ nT, $1.5$ nT, and we assume that the electromagnetic interaction parameter $\zeta$ = $2\times10^{-5}$ A$\cdot$m is characterized by a current intensity $I$ = $1$ mA.
				(c), (d) The magnetic field $B=0$ nT, $0.01$ nT, $0.1$ nT with the current intensity $I=1$ mA. 
				(e), (f) The current intensity $I=1$ mA, $2.5$ mA, $5$ mA with the magnetic field is $B=0.1$ nT. 
				The other parameters are $\omega_{m}$ = $2\pi\times0.1$ MHz $\omega_{m}$ = $2\pi\times0.1$ MHz,
				$m$ = $100$ pg, $G$ = $-2\pi\times11$ MHz/nm, $\gamma_{m}$ = $2\pi\times0.2$ kHz, $\kappa_{a}$ = $0.2\omega_{m}$, $g_{a}$ = $-\kappa_{a}$, $P_{\text{d}}$ = $100$ pW, $J$ = $0$, $\Delta_{a}$ = $-\omega$, and the wavelength of the driving field $\lambda_{\text{d}}$ = $2\pi c/\omega_{\text{d}}$ = $532$ nm ($c$ represents the speed of light in vacuum).
			}\label{Fig2}
		\end{figure}
		
		By using the input-output relation of cavity $a_{\text{out}}\left(t\right)=a_{\text{in}}\left(t\right)-\sqrt{\eta_{c}\kappa_{a}}a_{1}$, we obtain the output field			
		\begin{align}
			a_{\text{out}}\left(t\right)=&\left(\epsilon_{\text{d}}-\bar{a}_{1}\sqrt{\eta_{c}\kappa_{a}}\right)e^{-\mathrm{i}\omega_{\text{d}}t}+\left(\epsilon_{\text{p}}-A_{1}^{-}\sqrt{\eta_{c}\kappa_{a}}\right)e^{-\mathrm{i}\omega_{\text{p}}t}\nonumber\\
			&-A_{1}^{+}\sqrt{\eta_{c}\kappa_{a}}e^{-\mathrm{i}\left(2\omega_{\text{d}}-\omega_{\text{p}}\right)t}-A_{2}^{-}\sqrt{\eta_{c}\kappa_{a}}e^{-\mathrm{i}\left(2\omega_{\text{p}}-\omega_{\text{d}}\right)t}\nonumber\\
			&-A_{2}^{+}\sqrt{\eta_{c}\kappa_{a}}e^{-\mathrm{i}\left(3\omega_{\text{d}}-2\omega_{\text{p}}\right)t}.
		\end{align}
		The first two terms denote the output fields with the frequencies of $\omega_{\text{d}}$ and $\omega_{\text{p}}$.
		The third term describes the Stokes (down‑conversion) process.
		The last two terms represent the output fields at the frequencies of $\omega_{\text{d}}+2\Omega$ and $\omega_{\text{d}}-2\Omega$, and are associated with the generation of the optical second‑order upper and lower sidebands.
		
		The transmission of the probe field and the efficiency of the $n$‑th order upper sideband are defined respectively as
		\begin{align}
			&t_{p}=1-\frac{A_{1}^{-}}{\epsilon_{\text{p}}}\sqrt{\eta_{c}\kappa_{a}},
			&\theta_{n}=\left|-\frac{\sqrt{\eta_{c}\kappa_{a}}}{\epsilon_{\text{p}}}A_{n}^{-}\right|.
		\end{align}
		
		The associated group delay for the probe field or the second-order upper sideband is given by~\cite{jiao_optomechanical_2018,safavi-naeini_electromagnetically_2011}:
		\begin{align}
			&\tau_{1}=\left.\frac{d\arg\left(t_{p}\right)}{d\Omega}\right|_{\Omega=\omega},
			&\tau_{2}=\left.\frac{d\arg\left(A_{2}^{-}\right)}{2d\Omega}\right|_{\Omega=\omega},
		\end{align}
		where $\text{arg}(\cdot)$ represents the argument of a complex number.
	
	\section{RESULTS AND DISCUSSION}\label{Sec.III}
		\subsection{Dependence of the second-order sideband on multiple parameters}
			
			\begin{figure}
				\centering
				\begin{minipage}{0.49\linewidth}
					\centering				
					\begin{overpic}[width=\linewidth]{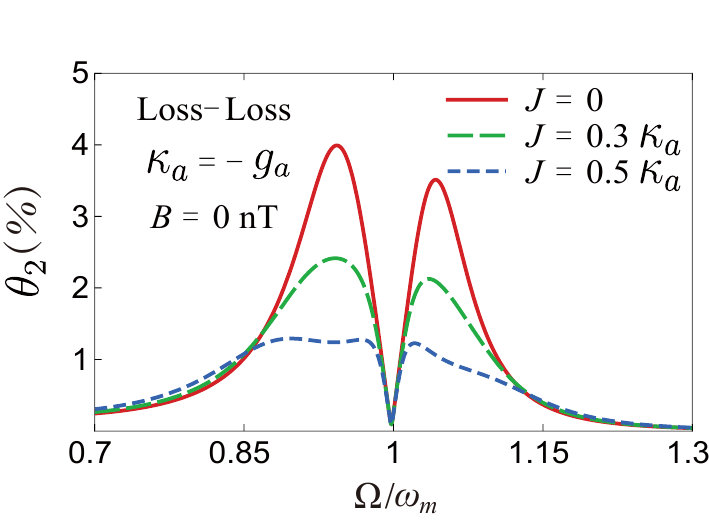}
						\put(0,70){(a)}
					\end{overpic}
				\end{minipage}
				\begin{minipage}{0.49\linewidth}
					\centering
					\begin{overpic}[width=\linewidth]{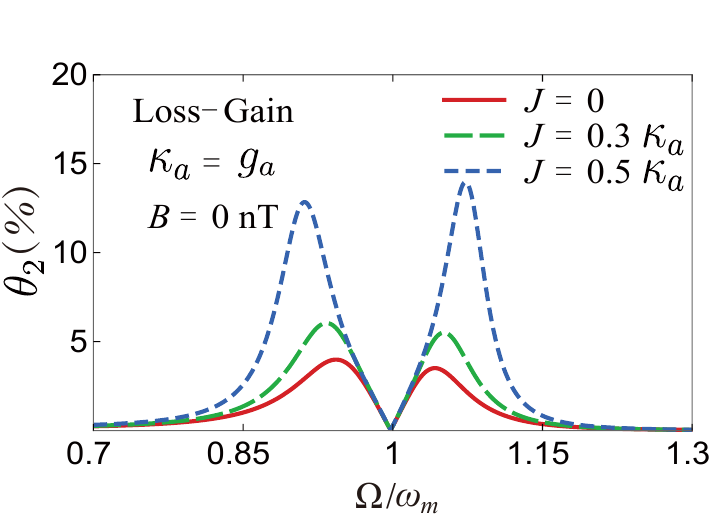}
						\put(0,70){(b)}
					\end{overpic}
				\end{minipage}
				\begin{minipage}{0.49\linewidth}
					\centering				
					\begin{overpic}[width=\linewidth]{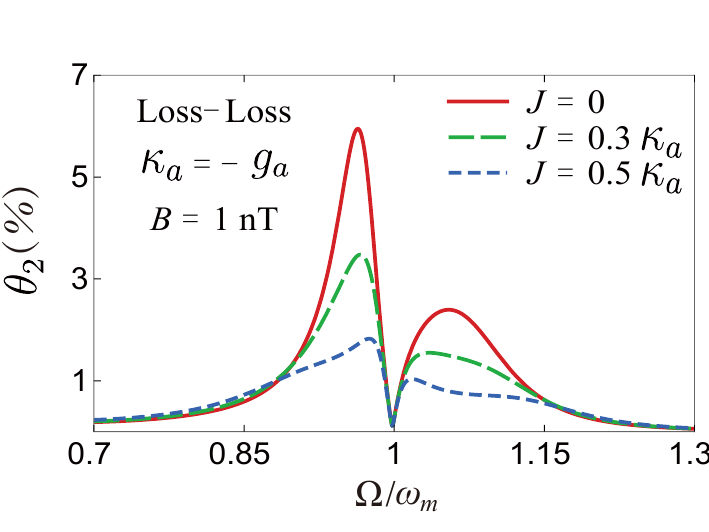}
						\put(0,70){(c)}
					\end{overpic}
				\end{minipage}
				\begin{minipage}{0.49\linewidth}
					\centering				
					\begin{overpic}[width=\linewidth]{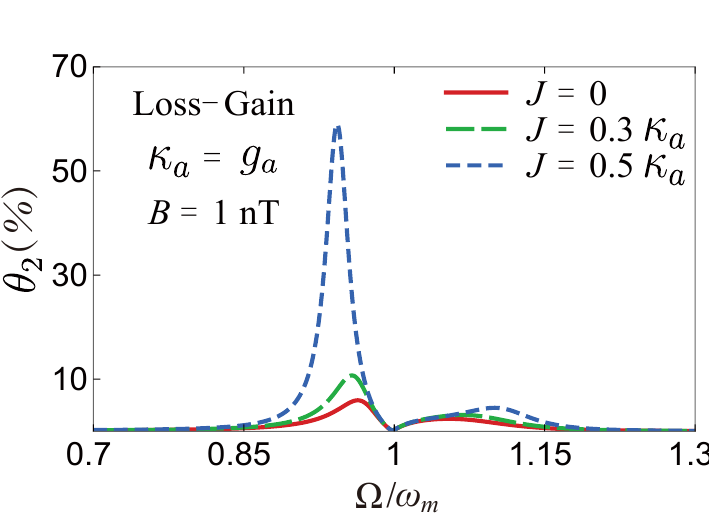}
						\put(0,70){(d)}
					\end{overpic}
				\end{minipage}		
				\caption{
					Second-order upper sideband efficiency $\theta_{2}$ versus normalized detuning $\Omega/\omega_{m}$ for different $\PT$-symmetric configurations, photon tunneling strengths $J$, and magnetic fields $B$.
					(a) $B$ = $0$ nT, $\kappa_{a} = -g_{a}$, (b) $B$ = $0$ nT, $\kappa_{a} = g_{a}$, (c) $B$ = $1$ nT, $\kappa_{a} = -g_{a}$, (d) $B$ = $1$ nT, $\kappa_{a} = g_{a}$.
					The driving power is $P_{\text{d}}$ = $1$ nW, and other parameters are the same as those in Fig.~\ref{Fig2}(a).
				}\label{Fig3}
			\end{figure}
			Our analysis begins with the impact of the magnetic field $B$, which enters the dynamics much like an effective nonlinear gain~\cite{liao_generation_2023}, breaking the symmetry of the system.
			For the numerical study we adopt parameters aligned with prior experiments~\cite{liu_precision_2018,zhang_pt_2020,liu_proposed_2017,thompson_strong_2008}:
			resonance frequency of the mechanical oscillator $\omega_{m}$ = $2\pi\times0.1$ MHz, mass $m$ = $100$ pg, mechanical damping rate $\gamma_{m}$ = $2\pi\times0.2$ kHz, optomechanical coupling strength $G$ = $-2\pi\times11$ MHz/nm, loss cavity decay rate $\kappa_{a}$ = $0.2\omega_{m}$, driving wavelength $\lambda_{\text{d}}$ = $2\pi c/\omega_{\text{d}}$ = $532$ nm ($c$ represents the speed of light in vacuum, $\omega_{\text{d}}$ is the frequency of the driving field).
			In a standard OMIT process, a strong control field and a weak probe field generate an anti-Stokes field, creating a transparency window in the probe transmission spectrum~\cite{weis_optomechanically_2010}.
			Figures~\ref{Fig2}(a) and~\ref{Fig2}(b) show the variation of OMIT transmission window and the second-order sideband efficiency with the magnetic field. 
			As the applied magnetic field increases, the symmetry of both the transmission and the second-order sideband efficiency is broken.
			On either side of the symmetry axis, the transmission windows shift toward higher frequencies [Fig.~\ref{Fig2}(a)], while the peak difference in the second-order sideband efficiency becomes more pronounced [Fig.~\ref{Fig2}(b)].
			Consequently, this mechanism naturally leads to a method for measuring the external magnetic field, the details of which form the focus of the next section.
			
			The second-order sideband efficiency, shown in Figs.~\ref{Fig2}(c) and~\ref{Fig2}(d), demonstrates higher sensitivity to the magnetic field than its first-order counterpart, allowing for the discrimination of fields as weak as $0.1$ nT and $0.01$ nT.
			Since the electromagnetic coupling strength $\zeta$ is proportional to the current intensity $I$, the current plays a role analogous to that of the magnetic field and could enhance the precision of magnetic field measurements [Figs.~\ref{Fig2}(e) and~\ref{Fig2}(f)].

			The dependence of second-order sideband generation on the $\PT$-symmetric structure, photon tunneling strength $J$, and magnetic field $B$ is examined herein, as illustrated in Fig~\ref{Fig3}.
			When no magnetic field is initially present [Figs.~\ref{Fig3}(a) and~\ref{Fig3}(b)], increasing the photon tunneling strength $J$ reduces the second‑order sideband in the Loss‑Loss system but enhances it in the Loss‑Gain system.
			The Loss‑Loss system operates in the $\PT$‑symmetry‑broken phase under the same given parameters, so strong tunneling suppresses sideband generation and results in a smaller sideband amplitude relative to the Loss‑Gain system.
			In contrast, in the Loss-Gain system, the active gain in the auxiliary cavity compensates for the loss of the lossy cavity~\cite{xiao_tunable_2020}.
			This means that the effective amplification of second-order sideband signals can be achieved by virtue of the synergistic effect of the photon tunneling strength $J$ and the $\PT$-symmetric structure.
			When a magnetic field is considered, as shown in Figs.~\ref{Fig3}(c) and~\ref{Fig3}(d), the combined action of the magnetic field and the tunneling strength within the $\PT$‑symmetric structure allows the second‑order sideband amplitude to increase from 3.99\% to 58.9\% [an enhancement of nearly 15 times compared with Fig.~\ref{Fig3}(a)].
	
				\begin{figure}
				\centering
				\begin{minipage}{0.49\linewidth}
					\centering				
					\begin{overpic}[width=\linewidth]{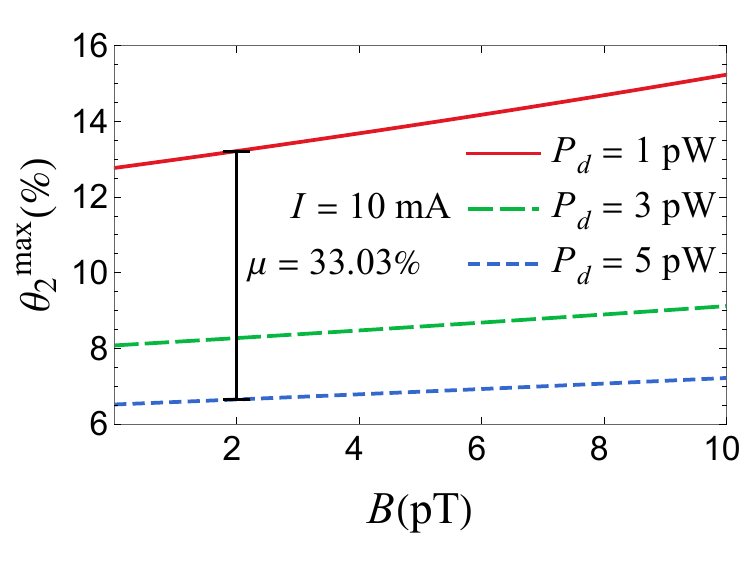}
						\put(0,72){(a)}
					\end{overpic}
				\end{minipage}
				\begin{minipage}{0.49\linewidth}
					\centering
					\begin{overpic}[width=\linewidth]{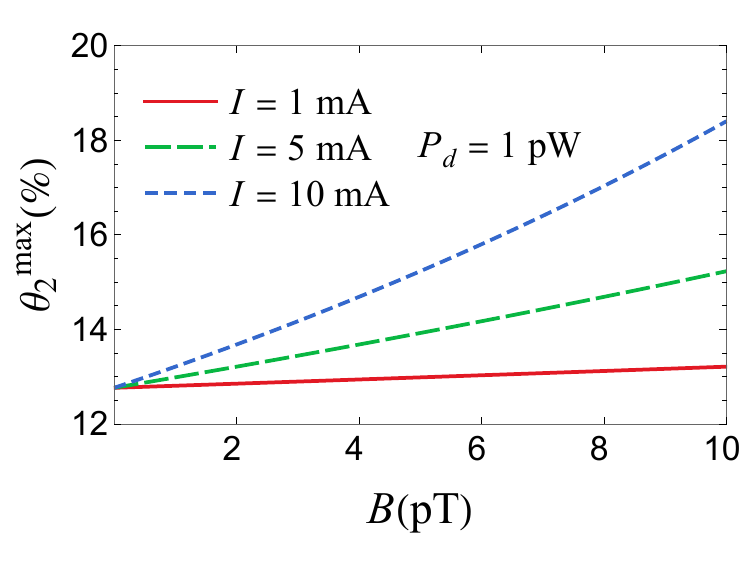}
						\put(0,72){(b)}
					\end{overpic}
				\end{minipage}\\		
				\caption{
					Maximum value $\theta_{2}^{\text{max}}$ as a function of magnetic field $B$ for (a) different driving powers $P_{d}$ and $I$ = $10$ mA, (b) different current intensities $I$ and $P_{\text{d}}$ = $1$ pW. The detuning, photon tunneling strength, gain rate are $\Delta_{a}$ = $0$, $J$ = $0.5\kappa_{a}$, $g_{a}$ = $\kappa_{a}$, respectively, and other parameters are the same as those in Fig.~\ref{Fig2}(a).
				}\label{Fig4}
			\end{figure}
			
		\subsection{Enhanced  magnetic field sensing via second-order sidebands in $\PT$-symmetric cavity optomechanics}
			The above analysis has demonstrated a significant enhancement of second-order sidebands in $\PT$-symmetric COM.
			In this section, we investigate the accuracy of magnetic field detection utilizing the enhanced second-order sidebands.
			When the two optical cavities exhibit balanced loss and gain, i.e., $g_{a}$ = $-\kappa_{a}$, the system reaches a loss-gain equilibrium.
			Through coupling between the cavities, two supermodes are formed, whose eigenfrequencies are given by $\omega_{\pm} = \omega_{a} \pm (J^{2} - \kappa_{a}^{2}/4)^{1/2}$.
			At $J$ = $0.5\kappa_{a}$, the eigenvalues become degenerate, corresponding to a $\PT$-symmetry phase transition~\cite{lü_symmetrybreaking_2015,konotop_nonlinear_2016}.
			Previous work has established that the system exhibits enhanced sensitivity to magnetic field variations near this $\PT$-symmetry breaking point~\cite{zhang_pt_2020}.
			
			We therefore study the variation of the maximum second‑order sideband efficiency $\theta_{2}^{\text{max}}$ versus magnetic field $B$ under this condition ($J$ = $0.5\kappa_{a}$), as shown in Fig.~\ref{Fig4}. 
			As illustrated in Fig.~\ref{Fig4}(a), reducing the driving power $P_{\text{d}}$ of the lossy cavity effectively prevents the electromagnetic interaction from being masked by the optomechanical interaction.
			This leads to a clear signal resolution at a magnetic field, e.g., at $B=2$ pT with the sensitivity coefficient $\mu\coloneq |\theta_{2}(B_1)-\theta_{2}(B_2)|/[\theta_{2}(B_1)+\theta_{2}(B_2)]\times100\%=33.03\%$.
			Alternatively, increasing the current intensity $I$ enhances the coupling between the magnetic field and the mechanical oscillator, thereby improving the magnetic field sensitivity, as shown in Fig.~\ref{Fig4}(b).
			Consequently, regulating the driving power and current provides a viable approach for picotesla-level magnetic field sensing based on the second-order sideband efficiency.
			
			The results above show that the second-order sideband efficiency depends strongly on the magnetic field at the $\PT$-symmetry breaking point $J$ = $0.5\kappa_{a}$ and can be further enhanced by tuning the current intensity $I$ and the driving power $P_{\text{d}}$.
			The numerical results demonstrate a picotesla-level sensing sensitivity, which surpasses the performance of first-order sideband-based schemes by an order of magnitude~\cite{zhang_pt_2020}.
			Compared with traditional magnetometry approaches, the present scheme, based on second-order sideband generation in $\PT$‑symmetric COM, offers several key advantages.
			First, the combination of sensitive nonlinear optomechanical interactions and the exceptional point of the $\PT$‑symmetric structure significantly enhances measurement accuracy.
			Second, while first-order sidebands typically require the driving power to exceed a threshold, second-order sidebands can be generated even under low driving conditions~\cite{xiong_higherorder_2012}.
			Third, COM are naturally suited for sensing weak magnetic field~\cite{liu_proposed_2017}, forces~\cite{børkje_signatures_2013} and other perturbations~\cite{liu_quadrature_2018,wang_optomechanical_2024} owing to their ultra‑low mass, high mechanical quality factors, and integrability~\cite{liu_precision_2018,xie_phonon_2017}.
			Detecting magnetic fields through the second-order sideband efficiency in transmission spectra is therefore experimentally feasible within existing optomechanical platforms~\cite{liu_precision_2018}.
			
			\begin{figure}
				\centering
				\begin{minipage}{0.49\linewidth}
					\centering				
					\begin{overpic}[width=\linewidth]{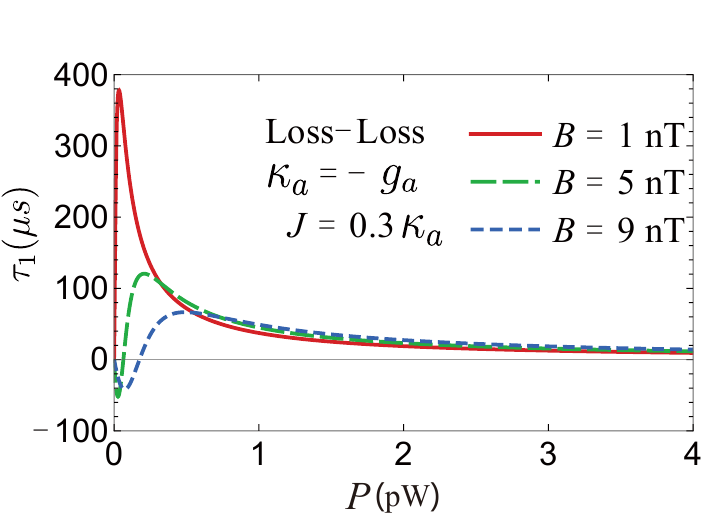}
						\put(0,67){(a)}
					\end{overpic}
				\end{minipage}
				\begin{minipage}{0.49\linewidth}
					\centering
					\begin{overpic}[width=\linewidth]{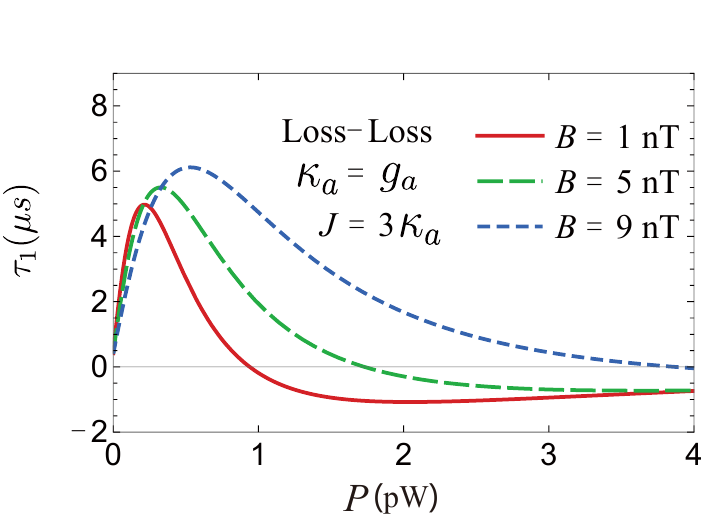}
						\put(0,67){(b)}
					\end{overpic}
				\end{minipage}
				\begin{minipage}{0.49\linewidth}
					\centering				
					\begin{overpic}[width=\linewidth]{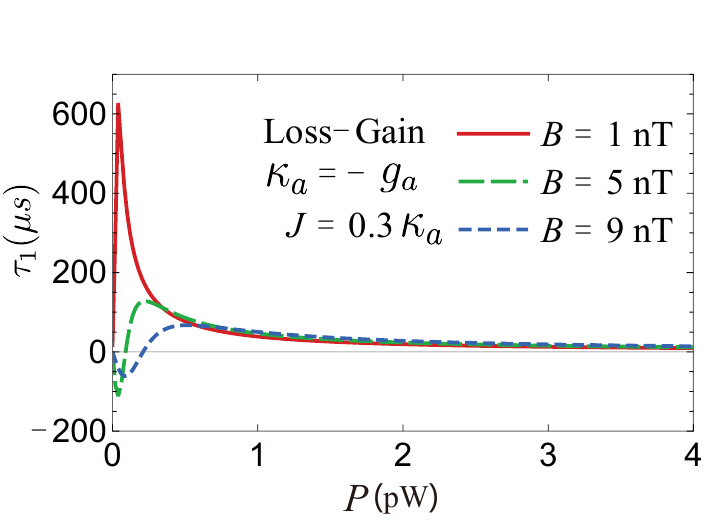}
						\put(0,67){(c)}
					\end{overpic}
				\end{minipage}
				\begin{minipage}{0.49\linewidth}
					\centering				
					\begin{overpic}[width=\linewidth]{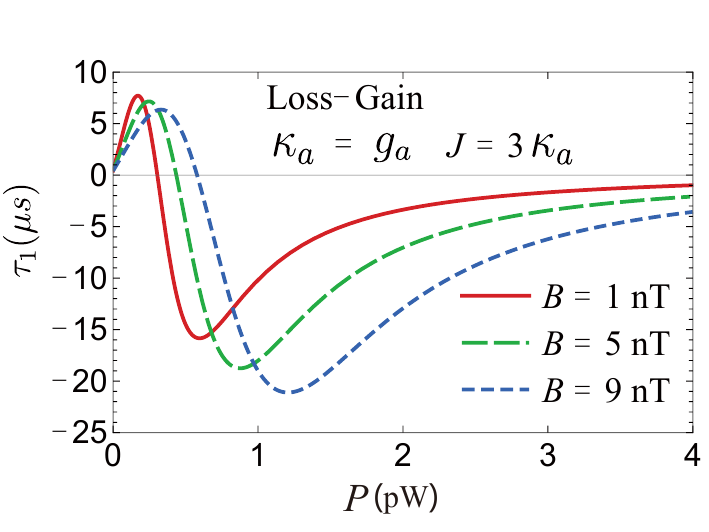}
						\put(0,67){(d)}
					\end{overpic}
				\end{minipage}		
				\caption{Group delay $\tau_{1}$ of probe light versus the driving power $P_{d}$ under various magnetic fields $B$.
					(a), (b) Loss-Loss cavity $\kappa_{a}$ = $-g_{a}$ with $J=0.3\kappa_{a}$,  and $J=3\kappa_{a}$, respectively.
					(c), (d) Loss-Gain cavity $\kappa_{a}$ = $g_{a}$ with $J=0.3\kappa_{a}$,  and $J=3\kappa_{a}$, respectively.
					Other parameters are the same as those in Fig.~\ref{Fig2}(a).
				}\label{Fig5}
			\end{figure}
		\subsection{Mechanism for controlling group delay and slow-fast light in $\PT$-symmetric cavity optomechanics}
\begin{figure}
				\centering
				\begin{minipage}{0.49\linewidth}
					\centering				
					\begin{overpic}[width=\linewidth]{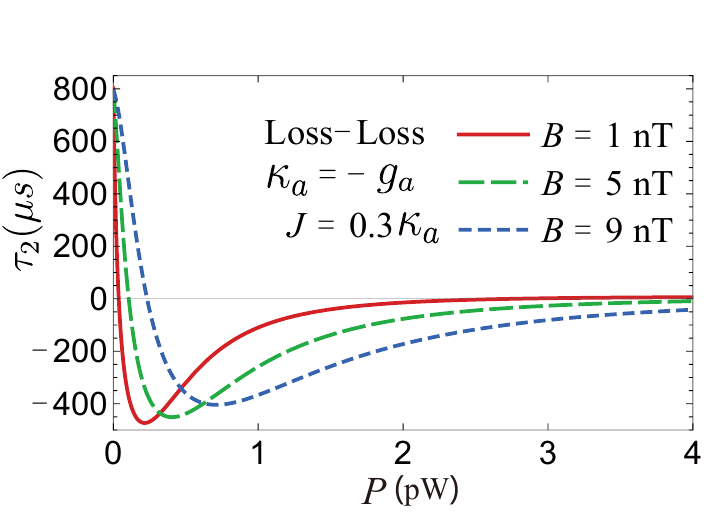}
						\put(0,70){(a)}
					\end{overpic}
				\end{minipage}
				\begin{minipage}{0.49\linewidth}
					\centering
					\begin{overpic}[width=\linewidth]{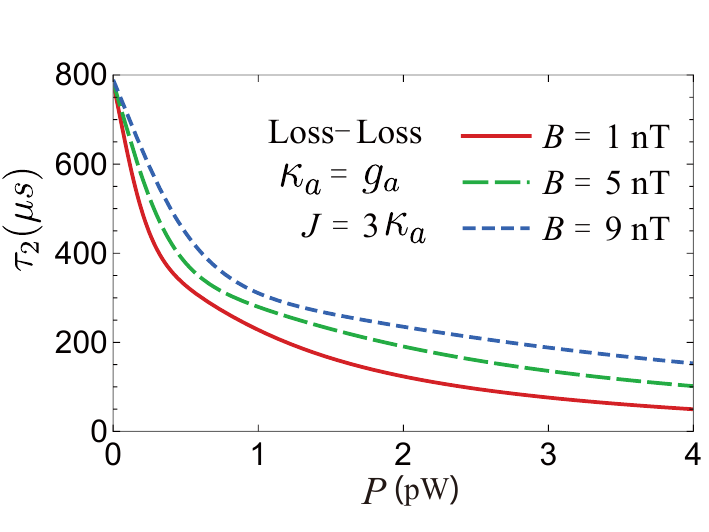}
						\put(0,70){(b)}
					\end{overpic}
				\end{minipage}
				\begin{minipage}{0.49\linewidth}
					\centering				
					\begin{overpic}[width=\linewidth]{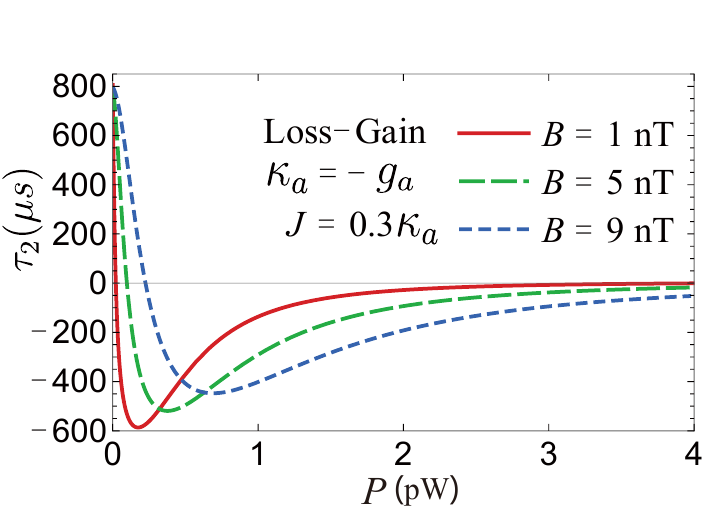}
						\put(0,70){(c)}
					\end{overpic}
				\end{minipage}
				\begin{minipage}{0.49\linewidth}
					\centering				
					\begin{overpic}[width=\linewidth]{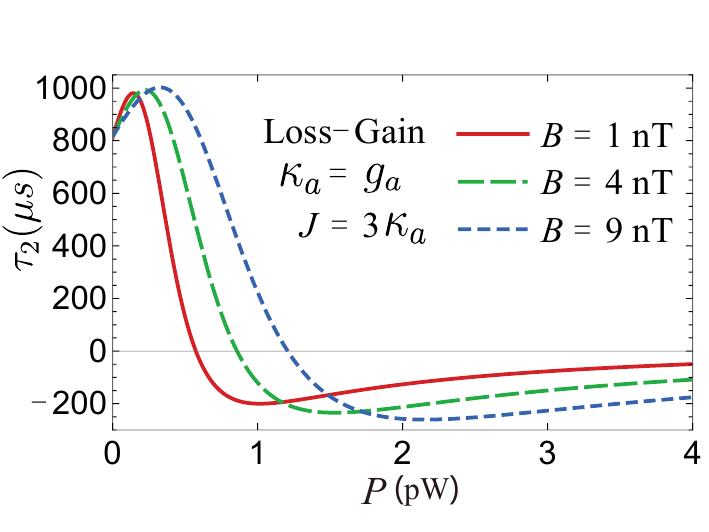}
						\put(0,70){(d)}
					\end{overpic}
				\end{minipage}		
				\caption{
					Group delay  $\tau_{2}$ of second-order upper sideband versus the driving power $P_{\text{d}}$ under various magnetic fields $B$.
					(a), (b) Loss-Loss cavity $\kappa_{a}$ = $-g_{a}$ with $J=0.3\kappa_{a}$,  and $J=3\kappa_{a}$, respectively.
					(c), (d) Loss-Gain cavity $\kappa_{a}$ = $g_{a}$ with $J=0.3\kappa_{a}$,  and $J=3\kappa_{a}$, respectively.
					Other parameters are the same as those in Fig.~\ref{Fig2}(a).
				}\label{Fig6}
			\end{figure}

				The group delay of light can be engineered in optomechanical systems.
			We investigate the fast- and slow-light effects, quantified by the group delays $\tau_{i}$ with the $i$-order sideband, in four system configurations (Loss-Loss and Loss-Gain, $\PT$-symmetry and broken $\PT$-symmetry) under an applied magnetic field, as seen in Figs.~\ref{Fig5} and~\ref{Fig6}.

			We present the dependence of the probe group delay  $\tau_{1}$ on the driving power $P_{d}$ under various magnetic fields $B$, as shown in Fig~\ref{Fig5}.
			At weak tunneling strength [Figs.~\ref{Fig5}(a) and~\ref{Fig5}(c)], slow light is dominant ($\tau_{1}>0$).
			Here, the magnetic field reduces both the positive peak of the probe group delay and widens the driving power window where fast light appears.
			As tunneling strength increases, fast light gradually becomes dominant ($\tau_{1}<0$), with a more pronounced alternation between slow and fast light [Figs.~\ref{Fig5}(b) and~\ref{Fig5}(d)].
			In this regime, the positive and negative peaks of $\tau_{1}$ decrease markedly, while the magnetic field extends the driving power range of slow light.
			Comparing Figs.~\ref{Fig5}(c) and~\ref{Fig5}(d) with Figs.~\ref{Fig5}(a) and~\ref{Fig5}(b) respectively, tuning the cavities to the loss‑gain equilibrium condition increases the positive and negative peaks and enhances the fast‑light effect.
			
			In Fig.~\ref{Fig6}, we show the group delay $\tau_{2}$ of the second-order sideband as a function of the driving power $P_{\text{d}}$ under various magnetic fields $B$.
			Similarly to the group delay of the probe light, a comparison of Figs.~\ref{Fig6}(c) and~\ref{Fig6}(d) with Figs.~\ref{Fig6}(a) and~\ref{Fig6}(b) reveals that the loss‑gain equilibrium condition raises both positive and negative peaks of $\tau_{2}$ and enhances fast-light effect.
			However, increasing the magnetic field invariably broadens the driving-power range over which slow light persists.
			At weak tunneling strength, a stronger field reduces the negative peak of $\tau_{2}$ and shifts its occurrence to higher driving power.
			With larger tunneling strength, slow light becomes more dominant, e.g., in the Loss‑Loss cavity [Fig.~\ref{Fig6}(b)], slow light survives over a substantially wide power range.
			As seen in Fig.~\ref{Fig6}(d), increasing the magnetic field can simultaneously extend the slow‑light persistence range and elevate its peak.

			According to our results, both slow- and fast-light effects can be flexibly controlled via the magnetic field $B$, photon tunneling strength $J$, and loss-gain equilibrium condition ($\PT$‑symmetric structure).
			The $\PT$-symmetric structure enables slow‑fast switching via driving power and enhances group delay through magnetic and tunneling control, pushing delay times into the micro‑ to millisecond level.
			The attainment of longer group delays, together with operation over expanded driving power ranges, carries significant potential for applications in optical memory~\cite{lei_quantum_2023} and quantum communication~\cite{krauss_why_2008,boyd_controlling_2009,fukui_alloptical_2021}.
			
	\section{Conclusion}\label{Sec.IV}
		Second-order sideband generation and tunable slow-fast light effects in a $\PT$-symmetric electromagnetic hybrid COM system are explored, alongside the proposal of a corresponding magnetic-field sensing scheme.
		By characterizing the system dynamics via the semiclassical Heisenberg-Langevin formalism, we demonstrate that the generation efficiency of second-order sidebands can be significantly enhanced by tuning the magnetic field, photon tunneling strength, and $\PT$-symmetric configuration.
		The proposed magnetometer operates near the $\PT$-symmetry-breaking point and exploits these enhanced second-order sidebands.
		Compared with previous sensing schemes relying on first-order sidebands, our scheme exhibits superior performance, achieving a detection precision at the picotesla level.
		Furthermore, we investigate the group delay behavior of the probe light and second-order sidebands.
		It is shown that tuning the magnetic field, photon tunneling strength, and $\PT$-symmetric configuration not only extends the group delay into the millisecond level but also enables switching between slow light and fast light by regulating the driving power.
		Our results are expected to find applications in optical storage, quantum metrology, and quantum sensing.
	
	\section*{Acknowledgments}
		This work is supported by the National Natural Science Foundation of China (Grants No.~12505032 and No.~12247158), and the Fundamental Research Funds of Zhejiang University of Science and Technology (Grant No.~2025QN063). J.Y.L. acknowledges the support from the China CEEC Educational Exchange Program (Grant No.~2023269).

	\section*{Appendix}	
			Here, we present a detailed derivation of Eqs.~\eqref{A-1}--\eqref{A-2}.
		Following the assumptions introduced in the main text, we solve the evolution equations and derive the amplitude of the second-order sideband.
		Substituting the Ans\"atze into the evolution equations, we obtain the following twelve equations:		
		\begin{align}
			&\left(\Delta_{1}+\Omega\right)A_{1}^{-}=G\left(\bar{a}_{1}X_{1}+X_{1}^{*}A_{2}^{-}+X_{2}A_{1}^{+}\right)+JB_{1}^{-}+\mathrm{i}\sqrt{\eta_{c}\kappa_{a}}\epsilon_{\text{p}},\nonumber\\
			&\left(\Delta_{1}-\Omega\right)A_{1}^{+}=G\left(\bar{a}_{1}X_{1}^{*}+X_{1}A_{2}^{+}+X_{2}^{*}A_{1}^{-}\right)+JB_{1}^{+},\nonumber\\
			&\left(\Delta_{1}+2\Omega\right)A_{2}^{-}=G\left(\bar{a}_{1}X_{2}+X_{1}A_{1}^{-}\right)+JB_{2}^{-},\nonumber\\
			&\left(\Delta_{1}-2\Omega\right)A_{2}^{+}=G\left(\bar{a}_{1}X_{2}^{*}+X_{1}^{*}A_{1}^{+}\right)+JB_{2}^{+},\nonumber\\
			&\left(\Delta_{2}+\Omega\right)B_{1}^{-}=JA_{1}^{-},\nonumber\\
			&\left(\Delta_{2}-\Omega\right)B_{1}^{+}=JA_{1}^{+},\nonumber\\
			&\left(\Delta_{2}+2\Omega\right)B_{2}^{-}=JA_{2}^{-},\nonumber\\
			&\left(\Delta_{2}-2\Omega\right)B_{2}^{+}=JA_{2}^{+},\nonumber\\
			&\left(\omega^{2}\!-\!\Omega^{2}\!-\!\mathrm{i}\gamma_{m}\Omega\right)\!X_{1}\!=\!-\frac{\hbar G}{m}\left[\bar{a}_{1}\!\left(\!A_{1}^{+}\!\right)\!^{*}\!+\!\bar{a}_{1}^{*}A_{1}^{-}\!+\!\left(\!A_{1}^{-}\!\right)\!^{*}A_{2}^{-}\!+\!\left(\!A_{2}^{+}\!\right)\!^{*}A_{1}^{+}\right],\nonumber\\
			&\left(\omega^{2}\!-\!\Omega^{2}\!+\!\mathrm{i}\gamma_{m}\Omega\right)\!X_{1}^{*}\!=\!-\frac{\hbar G}{m}\left[\bar{a}_{1}\!\left(\!A_{1}^{-}\!\right)\!^{*}\!+\!\bar{a}_{1}^{*}A_{1}^{+}\!+\!\left(\!A_{2}^{-}\!\right)\!^{*}A_{1}^{-}\!+\!\left(\!A_{1}^{+}\!\right)\!^{*}A_{2}^{+}\right],\nonumber\\
			&\left(\omega^{2}\!-\!4\Omega^{2}\!-\!2\mathrm{i}\gamma_{m}\Omega\right)\!X_{2}\!=\!-\frac{\!\hbar G\!}{m}\!\left[\bar{a}_{1}\!\left(\!A_{2}^{+}\!\right)\!^{*}\!+\!\bar{a}_{1}^{*}A_{2}^{-}\!+\!\left(\!A_{1}^{+}\!\right)\!^{*}A_{1}^{-}\right],\nonumber\\
			&\left(\omega^{2}\!-\!4\Omega^{2}\!+\!2\mathrm{i}\gamma_{m}\Omega\right)\!X_{2}^{*}\!=\!-\frac{\!\hbar G\!}{m}\!\left[\bar{a}_{1}\!\left(\!A_{2}^{-}\!\right)\!^{*}\!+\!\bar{a}_{1}^{*}A_{2}^{+}\!+\!\left(\!A_{1}^{-}\!\right)\!^{*}A_{1}^{+}\right],\tag{A1}
		\end{align}
		where $\Delta_{1}\coloneqq\Delta+\mathrm{i}\kappa_{a}/2=(\Delta_{a}-G\bar{x})+\mathrm{i}\kappa_{a}/2$.				
		The second-order sideband is treated as a second-order process whose amplitude is significantly smaller than that of the probe field.
		Under this approximation, the preceding equations can be simplified into the following two sets: one describing the first-order sideband,
		\begin{align}
			&\!\left(\!\Delta_{1}\!-\!\Omega\!-\!\frac{J^{2}}{\Delta_{2}\!-\!\Omega}\right)\!A_{1}^{+}\!=\!G\left(\bar{a}_{1}X_{1}^{*}\!+\!X_{1}A_{2}^{+}\!+\!X_{2}^{*}A_{1}^{-}\right),\nonumber\\
			&\!\left(\!\Delta_{1}\!+\!\Omega\!-\!\frac{J^{2}}{\Delta_{2}\!+\!\Omega}\right)\!A_{1}^{-}\!=\!G\left(\bar{a}_{1}X_{1}\!+\!X_{1}^{*}A_{2}^{-}\!+\!X_{2}A_{1}^{+}\right)\!+\!\mathrm{i}\sqrt{\eta_{c}\kappa_{a}}\epsilon_{\text{p}},\nonumber\\
			&\!\left(\!\omega^{2\!}\!-\!\Omega^{2}\!-\!\mathrm{i}\gamma_{m}\Omega\!\right)\!X_{\!1\!}\!=\!-\frac{\!\hbar G\!}{m}\left[\bar{a}_{1\!}\!\left(A_{1}^{+}\!\right)\!^{\!*\!}\!\!+\!\bar{a}_{1\!}^{*\!}A_{1\!}^{\!-\!}\!\!+\!\left(A_{1\!}^{-}\!\right)\!^{\!*}\!A_{2\!}^{\!-\!}\!\!+\!\left(A_{2}^{+}\!\right)\!^{*\!}\!A_{1}^{+}\!\right],\!\!\tag{A2}
		\end{align}				
		and the other for the second-order sideband,		
		\begin{align}
			&\left(\!\Delta_{1}\!+\!2\Omega\!-\!\frac{J^{2}}{\Delta_{2}\!+\!2\Omega}\!\right)A_{2}^{-}\!=\!G\left(\bar{a}_{1}X_{2}+X_{1}A_{1}^{-}\right),\nonumber\\
			&\left(\!\Delta_{1}\!-\!2\Omega\!-\!\frac{J^{2}}{\Delta_{2}\!-\!2\Omega}\!\right)A_{2}^{+}\!=\!G\left(\bar{a}_{1}X_{2}^{*}+X_{1}^{*}A_{1}^{+}\right),\nonumber\\
			&\left(\omega^{2}\!-\!4\Omega^{2}\!-\!2\mathrm{i}\gamma_{m}\Omega\right)X_{2}\!=\!-\frac{\hbar G}{m}\!\left[\bar{a}_{1}\!\left(\!A_{2}^{+}\!\right)\!^{*}\!+\!\bar{a}_{1}^{*}A_{2}^{-}\!+\!\left(\!A_{1}^{+}\!\right)\!^{*}\!A_{1}^{-}\right].\!\tag{A3}
		\end{align}			
		Finally, we obtain the amplitudes of the first-order upper (lower) sideband and the second-order upper sideband [presented as Eqs.~\eqref{A-1}--\eqref{A-2} in the main text],
		\begin{align}
			A_{1}^{-}&=\!\frac{\mathrm{i}\sqrt{\eta_{c}\kappa_{a}}\epsilon_{\text{p}}\left[h_{2}\left(\Omega\right)+h_{3}\left(\Omega\right)f\left(\Omega\right)\right]}{h_{1}\left(\Omega\right)},\tag{A4}\\
			(A_{1}^{+})^{*}&=\!\frac{-\mathrm{i}\hbar G^{2}\bar{a}_{1}^{*2}\sqrt{\eta_{c}\kappa_{a}}\epsilon_{\text{p}}\chi\left(\Omega\right)h_{3}\left(\Omega\right)}{h_{1}\left(\Omega\right)},\tag{A5}\\
			X_{1}&=\!\frac{-\mathrm{i}\hbar G\bar{a}_{1}^{*}\sqrt{\eta_{c}\kappa_{a}}\epsilon_{\text{p}}\chi\left(\Omega\right)h_{2}\left(\Omega\right)}{h_{1}\left(\Omega\right)},\tag{A6}\\
			A_{2}^{-}\!&=\!\frac{\!Gh_{2}\!\left(\!2\Omega\!\right)\!A\!_{1}\!^{-}\!X_{1}\!\lambda_{1}\!\left(\!2\Omega\!\right)\!-\!\lambda_{2}h_{3}\!^{2}\!\left(\!2\Omega\!\right)\!\chi\!\left(\!2\Omega\!\right)\!X_{1}\!^{2}}{h_{1}\!\left(2\Omega\right)h_{2}\!\left(2\Omega\right)},\tag{A7}
		\end{align}									
		where $h_{1}\left(\Omega\right)=[\Delta_{a}^{2}-\left(\mathrm{i}g_{a}/2-\Omega\right)^{2}][\Delta^{2}-(\mathrm{i}\kappa_{a}/2+\Omega)^{2}+2\Delta f(\Omega)]-2J^{2}[\Delta_{a}f(\Omega)+\Delta\Delta_{a}-(\mathrm{i}g_{a}/2-\Omega)(\mathrm{i}\kappa_{a}/2+\Omega)]+J^{4}$, $h_{2}(\Omega)=(\Delta_{a}-\mathrm{i}g_{a}/2+\Omega)[(\Delta_{a}+\mathrm{i}g_{a}/2-\Omega)(\Delta-\mathrm{i}\kappa_{a}/2-\Omega)-J^{2}]$, $h_{3}(\Omega)=(\Delta_{a}-\mathrm{i}g_{a}/2+\Omega)(\Delta_{a}+\mathrm{i}g_{a}/2-\Omega)$, $f(\Omega)=\hbar G^{2}|\bar{a}_{1}|^{2}\chi(\Omega)$, $\chi(\Omega)=1/[m(\omega^{2}-\Omega^{2}-\mathrm{i}\gamma_{m}\Omega))]$,  $\lambda_{1}(2\Omega)=[f(2\Omega)h_{3}(2\Omega)+h_{2}(2\Omega)]-\hbar\bar{a}_{1}^{2}G^{2}h_{3}(2\Omega)\chi(2\Omega)$, $\lambda_{2}=\hbar G^{4}\left|\bar{a}_{1}\right|^{2}\bar{a}_{1}^{*}$.	
					
	%

\end{document}